\documentclass[twocolumn]{aastex61}

\newcommand{\Msol}{\mathrm{M}_\odot}
\newcommand{\Gaia}{\textit{Gaia}~}

\usepackage{amsmath}
\usepackage{ulem}
\usepackage{color}
\usepackage{seqsplit}
\usepackage{hyperref}

\definecolor{blue}{rgb}{0.0,0.0,1.0}
\newcommand{\RP}{G_\mathrm{RP}}
\newcommand{\BP}{G_\mathrm{BP}}

\newcommand{\shao}{{Shanghai Astronomical Observatory, Chinese Academy of     
Sciences, No. 80 Nandan Road, Shanghai 200030, P. R. China}}

\submitjournal{ApJS}

\begin{document}


\title{A catalog of newly identified star
clusters in \textit{GAIA} DR2}
\author{Lei Liu}
\affiliation{\shao}
\author{Xiaoying Pang}
\affiliation{Xi'an Jiaotong-Liverpool University, 111 Ren'ai Road, Dushu Lake Science 
        and Education Innovation District, Suzhou 215123, Jiangsu Province, P. R. China. Xiaoying.Pang@xjtlu.edu.cn}
\affiliation{Shanghai Institute of Technology, 100 Haiquan Road,
        Fengxian district, Shanghai 201418, P.R. China}
\affiliation{Shanghai Key Laboratory for Astrophysics, Shanghai Normal University, 
        100 Guilin Road, Shanghai 200234, P.R. China}

\correspondingauthor{Xiaoying Pang}
\email{Xiaoying.Pang@xjtlu.edu.cn}

\begin{abstract}
We present the Star cluster Hunting Pipeline (SHiP) which can identify star 
clusters in \Gaia DR2 data, and establish a star cluster catalog for the 
Galactic disk. A Friend of Friend based cluster finder method is used to identify star 
clusters using 5-dimensional stellar parameters, $l, b, \varpi, 
\mu_\alpha\cos\delta$, and $\mu_\delta$. Our new catalog contains 2443 star 
cluster candidates identified from disk stars located within $|b|=25^\circ$ and 
with $G<18$\,mag. An automatic isochrone fitting scheme is applied to all 
cluster candidates.
With a combination of parameters obtained from isochrone fitting, 
we classify cluster candidates into three  classes (Class 1, 2 and 3). Class 1
clusters are the most probable star cluster candidates with the most  stringent criteria.    Most of these clusters are nearby  (within 4\,kpc).  
Our catalog is cross-matched with three  Galactic star cluster catalogs, 
\citet{K13}, \citet{CG18, CG19}, and \citet{Bica2019}. The proper motion and 
parallax of matched star clusters are in good agreement with these earlier catalogs. 
We discover 76 new star cluster candidates that are not listed in these 3 
catalogs. The majority of these are clusters older than log(age/yr)~=~8.0, 
and are located in the inner disk with $|b|<5^\circ$. The recent discovery of new star 
clusters suggests  that current Galactic star cluster catalogs are still 
incomplete.   Among the Class 1 cluster candidates, we find 56 candidates for  star cluster groups.  The pipeline, the catalog and the member list containing all candidates star clusters   and star cluster groups  have been made publicly available. 
\end{abstract}


\section{Introduction} \label{sec:intro}

Star clusters in the Galactic disk are important tracers of disk structure and 
dynamics. Open clusters 
(OCs) are distributed in the disk and are (mostly) young ($\le300$~Myr) and 
low-mass \citep[$<10^3~\Msol$; ][]{Dias2002, Piskunov2008}.
Perturbed by disk shocks, spiral arm passages, and encounters with  molecular 
clouds \citep{spi58, kru12}, OCs expand and disrupt at a timescale of 
200\,Myr~-~1\,Gyr \citep{yan13, Pang2018}. Therefore, only a small fraction of 
the observed OCs is older than 1\,Gyr \citep{K13}.

In order to study the formation and evolution of open clusters, much effort has 
been made to  compile OC catalogs, such as DAML02 \citep{Dias2002},  
MWSC \citep[Milky Way Star clusters; ][K13 hereafter]{K13} and \citet[][B19 
hereafter]{Bica2019}.
Based on the PPMXL proper motion catalog \citep{roe2010} and 2MASS photometry, 
MWSC developed an automated pipeline and identified  3006 star cluster objects. 
B19 made use of infrared photometry, and discovered several hundreds of new OC 
candidates in addition to those found by  \citet{K13}.
The second data release (DR2) of \textit{\Gaia} \citep{bro18}  revolutionizes 
OC studies by providing precise proper motions of individual stars, which are 
very suitable for  identifying star clusters in the multi-dimensional parameter 
space. Since its release in May 2018, several   groups  have 
published star cluster catalogs based on this archive.  
\citet{CG18} developed an unsupervised membership 
assignment code UPMASK to search for star clusters in \Gaia DR2. They managed to obtain parameters and members for 
  1229 star clusters (60 of these were new). Recently, \citet{Ginard2019} implemented a density based clustering algorithm, DBSCAN, 
and applied a supervised learning method \citep{Ginard2018} to \Gaia DR2 data.   53 new  OCs were detected in a region along the direction of Galactic anti-centre and the Perseus arm.  
  Their newly discovered OCs increase the number of known OCs in the direction of  Galactic anti-center by 22\% as compared to known OC populations  \citep[][CG18+19 hereafter]{K13,CG18,CG19}. Later on,  
\citet{CG19} applied a ``coarse-to-fine'' search method and again discovered 41 new star clusters in the direction of Perseus. Therefore, all these newly discovered star clusters point out a fact that many  more efforts are required to obtain a complete census of Galactic OCs.  
 In this study, we identify star clusters in \Gaia DR2 using our Star cluster Hunting 
Pipeline (SHiP). 
Star cluster candidates are first identified with   a   Friend of Friend (FoF) method.
This is directly inspired by the galaxy group finder algorithm first proposed by \citet{Yang2005}. There is a long tradition of using the FoF method in cosmological studies  \citep{Davis1985}. It has been successfully used for the identification of dark matter halos in cosmological simulations \citep{Springel2001} and galaxy groups in sky survey data \citep{Yang2007}. 
In this method, two particles are assigned to one group if their distance is smaller than  $b_\mathrm{FoF}$ (the linking length factor) times the mean separation of particles in the volume.
Furthermore, the FoF method is applicable to multi-dimensional parameter spaces and does not require any prior information.   At the same time, it is suitable for identifying star clusters using stellar positions, parallaxes and proper motions. Besides the FoF cluster finder, SHiP   employs  a full set of methods to further validate star clusters, including   an automatic  isochrone fitting   method  and a star cluster classification scheme based on a set of parameters   derived  during the fitting process.   The efficiency of SHiP is greatly aided by parallel computation.  At the same time, SHiP is 
capable of identifying star clusters from a huge amount of \Gaia DR2 data without 
requiring any star cluster catalog as input. 

This paper is organized as follows. The sample selection from \Gaia DR2 is presented in Sec.~\ref{sec:sample_select}. We introduce SHiP in Sec.~\ref{sec:iden_sc}, including the FoF-based star cluster finder (Sec.~\ref{sec:fof_sc}), the  isochrone fitting method (Sec.~\ref{sec:iso_fitting}) and the classification scheme (Sec.~\ref{sec:cls_sc}). In Sec.~\ref{sec:sc_property}, we describe the general properties of the star cluster candidates (Class 1, 2, and 3), the newly discovered star clusters,   and candidates of star cluster groups . Finally, a summary is presented in Sec.~\ref{sec:summary}. 

\begin{figure}
	\plotone{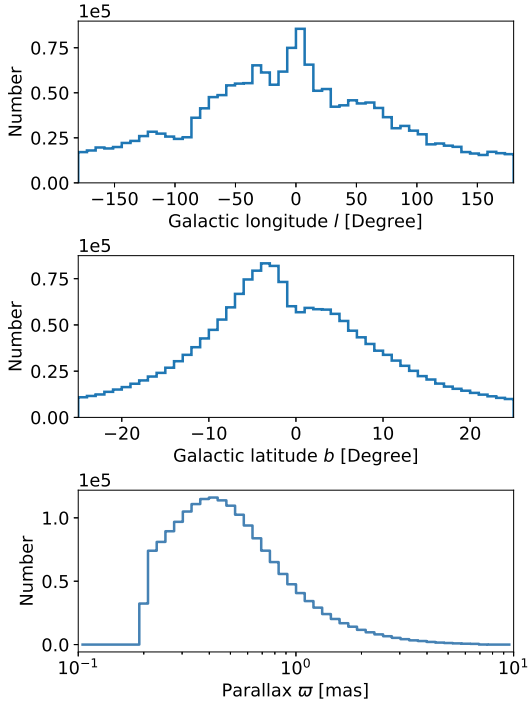}
	\caption{Histograms of $l$, $b$ and $\varpi$ of stars in the primary sample. \label{fig:starhist}}
\end{figure}

\section{Sample Selection of \Gaia DR2}
\label{sec:sample_select}

The \Gaia DR2 contains  over one billion sources,  most of which have  high-precision astrometric data including parallaxes ($\varpi$), proper motions  ($\mu_\alpha\cos\delta$ and $\mu_\delta$), and  photometry in three broad bands, the $G$ (330~--~1050 nm), $G_{\rm BP}$ (330~--~680~nm) and $G_{\rm RP}$ (630~--~1050~nm) bands. 
The median uncertainties of $\varpi$ and proper motions (without considering the systematic
errors) generally increase toward faint stars. The uncertainty in $\varpi$ is $\sim$~0.04~mas for sources with $G$\,$<$\,14~mag, 0.1~mas for $G$\,$\approx$\,17~mag, 
and $\sim$0.7~mas for $G$\,$\approx$\,20~mag,  corresponding to 0.05, 0.2 and 1.2\,mas/yr of uncertainty in proper motions, respectively
\citep{lin18}.

The majority of OCs are located near the Galactic plane with $b<20^\circ$ \citep{Dias2002,K13,Ginard2018}. Therefore,  we select stars up to $|b| = 25^\circ$ from \Gaia DR2 to identify disk star clusters. 
At the same time, in order to exclude observational artifacts due to faintness, we apply the following photometric and quality cuts:  
\begin{itemize}
\item[\textbullet] $G < 18$\,mag,
\item[\textbullet] $\mu_\alpha\cos\delta <$ 30~mas/yr, $\mu_{\delta} <$ 30~mas/yr,
\item[\textbullet] 0.2~mas $< \varpi <$ 7.0~mas.
\end{itemize}

  We apply the cut at $G<$~18\,mag,   which is  the same as the criterion used in Cantat-Gaudin et al. (2018).
Since the parallax error at $G=$~18\,mag is $\sim$0.2\,mas, we select stars with a parallax larger than 0.2\,mas, which constrains our sample to a maximum distance of 5 kpc. Besides that, there are not many stars beyond 7 mas (bottom panel in Fig.~\ref{fig:starhist}).  
In total, 180,490,541 stars are retrieved for the star cluster identification. We call this set of stars the primary sample. 
Coordinates of stars in the primary sample are transformed  from epoch 2015.5 (\Gaia DR2) to 
epoch 2000, with their corresponding proper motions. 
Galactic longitude ($l$) and latitude ($b$) are calculated in the reference of epoch 2000   for a convenient cross-match with earlier catalogs (see Section 4).  

In this study, we use 8 stellar parameters of stars from {\it \Gaia} DR2 in the primary sample:  $l$, $b$, $\varpi$, $\mu_\alpha\cos\delta$, and  $\mu_\delta$, and magnitudes in $G $, $G_{\rm BP}$ and $G_{\rm RP}$ bands for the following procedures. Histograms of $l$, $b$, $\varpi$ of the primary sample are shown in Fig.~\ref{fig:starhist}. Stars tend to concentrate towards the Galactic center (upper panel). The peak of stars in the southern Galactic hemisphere is due to the Sun's  northern position off the disk     \citep[middle panel of Fig.~\ref{fig:starhist},][]{Bonatto2006,Reid2014,Gravity2019}. Parallaxes of stars in the primary sample are mostly below 2~mas (bottom panel of Fig.~\ref{fig:starhist}). 

\section{Star Cluster Hunting Pipeline}\label{sec:iden_sc}

\subsection{FoF star cluster finder}\label{sec:fof_sc}

\subsubsection{Domain partition}\label{sec:partition}
  SHiP adopts an FoF cluster finder to identify star clusters in the Galactic field.
To facilitate this procedure,  
we divide the entire search volume into multiple domains.  This operation 
enables us to carry out the cluster identification process  
in parallel, which greatly enhances the computational  efficiency.

Stars are assigned to different domains according 
to their 3-D spatial coordinates $(l, b, \varpi)$. 
The specific partition strategy takes three considerations into account:
\begin{itemize}
\item[\textbullet] The size of the domain should not be too large, 
otherwise the FoF method cannot be used due to the large surface density
change of background stars in a domain from one side to the other side 
(see the upper and middle panels in Fig.~\ref{fig:starhist}). The number of stars 
also increases as the size becomes large, thus reducing  the efficiency in the calculations.
\item[\textbullet] The size of the domain should not be too small, 
otherwise there will be not enough stars to carry out the identification. 
The size of domain should be larger than the typical size of a 
star cluster, so as to accommodate at least one star cluster in the domain.   Besides that, we require the minimum size of the
domain larger than $\sigma_\varpi$ in parallax dimension. We set the minimum size of the domain in each dimension as
  $r_\mathrm{sc}$. Here 
$r_\mathrm{sc}=10$~pc is the typical scale of   a  star 
cluster \citep{PZ2010}; $\sigma_\varpi = 0.2~\mathrm{mas}$ is the 
uncertainty of parallax for  $G = 18$\,mag. 
\item[\textbullet] There should be sufficiently large overlapping regions between two adjacent 
domains along each dimension, so as to guarantee that the star cluster can be identified adequately even when it is located at the border of the domain.
\end{itemize}

According to the above three criteria, we adopt the following partitioning scheme:
\begin{itemize}
	\item[\textbullet]    The volume is split  along each dimension $(l, b,
	\varpi)$
	  in the following sequence: firstly the parallax $\varpi$, secondly the Galactic 
	latitude $b$ in the corresponding parallax range, and finally the Galactic longitude $l$ 
	in the corresponding   ranges of the  parallax and the  Galactic latitude. 
	\item[\textbullet]   A domain will be split recursively into equal parts, when it 
	fulfills the following two   criteria:  
    \begin{itemize}
    \item   The size is larger than $2 r_\mathrm{sc}$ in   dimensions of  the
    Galactic latitude and longitude, and is larger than $2\sigma_\varpi$ 
    in the parallax dimension. 
    \item   The number of stars $n$ in the domain is larger than both $n_0/2^{k_\mathrm{split}}$ and $n_\mathrm{total}/2^{k_{\mathrm{split}, \varpi}+k_{\mathrm{split}, b}+k_{\mathrm{split}, l}}$. 
    Here $n_0$ is the number of stars before   the split  in this dimension, 
    $n_\mathrm{total}$ is the total number of stars in the primary sample, and 
    $k_\mathrm{split}$ is the tentative   split  order, which corresponds to $k_{\mathrm{split}, \varpi}$, $k_{\mathrm{split}, b}$, and $k_{\mathrm{split}, l}$ in the corresponding dimensions. In   the  current
    implementation, they are set to be $k_{\mathrm{split}, \varpi}$=3, $k_{\mathrm{split}, b}$=3, and $k_{\mathrm{split}, l}$=6, respectively.  
    \end{itemize}
    This partitioning  process is somewhat similar   to   the tree method often used 
    in cosmological simulations  \citep{Springel2005}. After  carrying out the above
    scheme, the whole search volume is divided into 4311 domains. 
	\item[\textbullet] We select stars in each corresponding domain based on the 3-D spatial coordinates $(l, b, \varpi)$ of the stars.  
	The size of overlapping regions is  $\sigma_\varpi$ in the parallax dimension and $r_\mathrm{sc}$ in the $l$ and $b$ dimensions ($r_\mathrm{sc}$ is converted to the corresponding angle at the given distance). 
\end{itemize}

\subsubsection{Cluster identification with FoF}\label{sec:cluster_iden_fof}
We use the FoF method to identify  star clusters in the 5-D 
parameter space $\mathbf{X} = \{l, b, \varpi, \mu_\alpha\cos\delta, \mu_\delta\}$. A cluster is identified when the distance of one star to its nearest 
neighbor is smaller than the linking length factor $b_\mathrm{FoF}$ times the average distance in the 
domain. We normalize each of the parameters in the 5-D 
parameter space to the range (0, 1) so that it is scale-free. The weight of parameters
\begin{equation}
\mathbf{w} = (\cos b, 1, 0.5, 1, 1) / (0.2\cos b + 0.7), 
\end{equation}
is applied to the normalized parameters.  The first term, $\cos b$, is due to the  
contraction of $l$ at a given $b$ in spherical geometry. Since the 
uncertainty in the  parallax is larger than that of the other parameters, we set the weight of parallax to 0.5 to 
reduce its influence in the cluster identification.
For distance calculations  we use the $L^2$ norm (Euclidean norm). The normalization factor in the denominator guarantees that $\sum_{i=1}^5 w_i = 5$. The linking length is set to $r=b_\mathrm{FoF}/N_\mathrm{star}^{1/5}$. $b_\mathrm{FoF}$ is set to 0.2, which is commonly adopted in the dark matter halo identification of cosmological simulations \citep{Springel2001}. $N_\mathrm{star}$
is the number of stars in each domain. The choice of weight and linking length   together with the partitioning scheme (Sec.~\ref{sec:partition}) is somewhat arbitrary, which inevitably introduces some noise and uncertainty. Therefore, the nature of these star cluster candidates identified in this way should be 
further confirmed by isochrone fitting and cluster classification (Sec.~\ref{sec:iso_fitting} and Sec.~\ref{sec:cls_sc}).
We keep  star clusters with more than 50 member stars for further merging and validation. 
 4885  star clusters candidates are detected within the 4311 domains in the primary sample. 
Some star cluster candidates appear in more than one domain. If more than 50 percent of the members in such star clusters are identical, we merge these two clusters. This merging process reduces the number of star cluster candidates to 2443. 


\subsection{Isochrone fitting Scheme}\label{sec:iso_fitting}
We further confirm star cluster candidates (Sec.\ref{sec:cluster_iden_fof}) by fitting their color-magnitude diagrams (CMDs) with a set of isochrones of different   ages  and   metallicities. The reliability of star cluster candidate detections will be   assessed  according to the derived parameters during isochrone fitting. 

\subsubsection{The Padova Isochrones}\label{sec:padova}
The isochrones adopted in this work are obtained from the Padova 
database \citep{Marigo2017} of stellar evolutionary  tracks\footnote{\href{http://stev.oapd.inaf.it/cgi-bin/cmd_3.0}{http://stev.oapd.inaf.it/cgi-bin/cmd\_3.0}}. 
The \Gaia DR2 passband photometric system is taken from \citet{Evans2018}. We adopt a log-normal  initial mass function  \citep{Chabrier2001}. A series of isochrones are generated from $\log 
(t/\mathrm{yr})=6.6~\mathrm{to}~10.13$ at steps of $\Delta(\log t)=0.05$ for 
metallicities ranging from $\log(Z/\mathrm{Z}_\odot) = -2.0$ to 0.5 with steps of 0.25. 

\subsubsection{Parameters from isochrone fitting}\label{sec:iso_parameter}
  
 The key to reliable isochrone fitting is the   fitting  function, which determines fitting accuracy. 
At the same time, optimization is carried out to minimize the difference between the data and the fitting function by searching the parameter space, which includes age, metallicity, distance modulus and extinction. The form of the fitting function together with the optimization method determines the fitting speed. Several studies produced isochrone fitting pipelines based on the above idea. 
\citet{Perren2015} used the Bayesian approach and genetic 
algorithm to maximize the likelihood of the fitted parameters.
Another optimization function is the residual hyper-surface (the discrepancy between the observed and simulated Hess diagrams) used by 
\citet{Bonatto2019}, which minimizes the  function with simulated annealing.

Automated isochrone fitting of 2443 star cluster candidates requires an efficient fitting scheme, which at the same time provides good accuracy. To fulfill these requirements, we propose the following fitting function: 
\begin{equation}\label{eq:opt_func}
\bar{d^2}=\sum_{k=1}^{n}|\mathbf{x}_k-\mathbf{x}_{k, nn}|^2 / n.
\end{equation}
Here $\mathbf{x}_k=[G_k+\Delta_G, (\BP-\RP)_k + \Delta_{\BP-\RP}]$ is the position of the 
$k$-th star in the CMD with absolute magnitudes. 
$\mathbf{x}_{k,nn}$ is the $k$-th star's corresponding nearest 
neighboring point in the isochrone table. 
Four parameters are obtained from isochrone fitting: $\Delta_G$ (distance modulus in $G$ magnitude), $\Delta_{\BP-\RP}$ (color excess  $E(\BP-\RP$)), metallicity $Z$ and age $t$. 
  We minimize $\bar{d^2}$, the mean square distance between cluster stars and their closest neighboring points in the isochrone.  
This approach is easy to implement and sensitive to the discrepancy between 
isochrones and the actual data.
The nearest neighbor in the isochrone can be easily found with the $k$-D Tree method,
  which faciliates speeding up the fitting process of the 770 isochrones (each isochrone contains over 1500 points) for each star cluster
candidate. Besides, the optimization of $\bar{d^2}$ is easily carried out by the Nelder-Mead algorithm \citep{NM1965} provided by the ``scipy'' package.

  The main sequence (MS) is significantly broadened at the faint end, due to the uncertainty in brightness of faint stars at $G>\,$17\,mag ($\delta G\sim$0.072\,mag), which is three times larger than those with $G<\,$17\,mag ($\delta G\sim$0.023\,mag). Approximately 30\% of the members among our identified clusters are fainter than $G$\,=\,17\,mag, which greatly affects the quality of the isochrone fitting.  Therefore, we restrict the isochrone fitting to stars brighter than $G$\,=\,17~mag.  This treatment reduces the number of stars in the fitting by approximately 50 percent, but substantially improves the quality of the fits. 




\subsection{Star cluster classification}\label{sec:cls_sc}
In order to evaluate the reliability of star cluster candidates detections using the FoF
cluster finder, we carry out a classification based on parameters  obtained from  
isochrone fitting.
Because of observational uncertainties, there is no unique parameter that can represent the reliability of a star cluster detection accurately. Instead, a combination of several parameters can be used to minimize the influence of the  uncertainties. Therefore, we classify candidates based on the following parameters.  

\textbullet~$\bar{d^2}$: the average square of the distance between cluster stars and an isochrone. It can be used to estimate the fitting quality. A small value of $\bar{d^2}$ represents a good isochrone
fitting. 

\textbullet~$r_\mathrm{n}$: the narrowness of the MS in the 
CMD. This parameter is used to distinguish real star clusters from 
false detections. For example, a real cluster tends to show a narrow MS, while an artifact  might be very broad. The narrowness of the MS is defined as 
$r_\mathrm{n}=|v_1 / 
v_2|$. Here $v_1$ and $v_2$ are the two eigenvalues of the covariance matrix 
$\mathbf{M}$ 
of the distribution of stars in the CMD, with $|v_1| < |v_2|$. The covariance matrix 
$\mathbf{M}$ is defined as:
\begin{equation}
\mathbf{M} = \left(\begin{matrix}
\overline{x_i x_i} & \overline{x_i y_i} \\
\overline{x_i y_i} & \overline{y_i y_i} \\
\end{matrix}\right).
\end{equation}
Here $x_i = (\BP-\RP)_i - \overline{\BP-\RP}$, $y_i = G_i - \overline{G}$. According to 
the definition of $r_\mathrm{n}$, a smaller value corresponds to a narrower MS. 
  There is a certain degree of degeneration between $r_\mathrm{n}$ and $\bar{d^2}$. A star cluster with a good isochrone fitting  usually has small values of both $\bar{d^2}$ and $r_\mathrm{n}$. 

\textbullet~$n_{G<17}$: the number of bright stars with $G<17$\,mag included in the 
isochrone fitting. 
Since photometric uncertainty broadens the MS at the faint end, $n_{G<17}$ eliminates contamination and guarantees the quality of the selected star cluster candidates.   Inevitably, this parameter will make our results biased towards  relatively nearby clusters that contain a sufficient number of  bright stars for isochrone fitting.  

\begin{figure}
	\caption{The distribution of   isochrone fitted ages of star clusters. 
	The black dashed line represents the 5 Myr age cut. \label{fig:age_hist}}
	\plotone{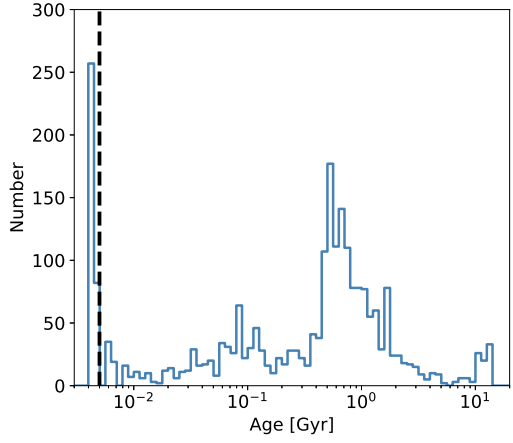}
\end{figure}

\textbullet~$t_\mathrm{age}$: the age of  star clusters derived from  isochrone 
fitting. The starting age of the Padova isochrones is $t=4$\,Myr ($10^{6.6}$\,yr). 
We plot the age distribution of star cluster candidates in Fig.~\ref{fig:age_hist}. The interpretation of the ages of the peak below 5~Myr (Fig.~\ref{fig:age_hist}) are  
highly uncertain since it approaches the lowest limit of the isochrone ages.
To ensure a reliable cluster classification, we consider only 
star cluster candidates 
with ages older than 5~Myr, which yield better fitting results. 

According to the above four criteria, we divide  the 2443 star cluster candidates into 3 classes:
\begin{itemize}
	\item Class 1: $n_{G<17}\ge 50$, $t_\mathrm{age} > 5~\mathrm{Myr}$, 
	$r_\mathrm{n} < 0.1$, $\bar{d^2} < 0.05$;
	\item Class 2: $n_{G<17}\ge 50$, $t_\mathrm{age} > 5~\mathrm{Myr}$, 
	$r_\mathrm{n} < 0.1$;
	\item Class 3: all other cases.
\end{itemize}
The numbers of star cluster candidates   in each group are: Class 1: 569 (23.3\%); Class 2: 127 (5.2\%); and Class 3: 1747 (71.5\%). 
   $n_{G<17}\ge 50$ restricts Class 1 and 2 to nearby star clusters,  mainly
within 4\,kpc.    The classification of 
star cluster candidates is influenced by the performance 
of our pipeline in the identification process, and should not be directly linked to  
the physical existence  of those star cluster candidates. According to our 
analysis,  members of Class 1 are likely star
clusters candidates. Class 2 and 3 are candidates that need  further confirmation. 
 However, genuine star 
clusters may exist among the members of Class 2 and Class 3 (see Sec.~\ref{sec:cross_matching}). General parameters for all 2443 candidates are  presented in Tab.~\ref{tab:cat_all}.  


\begin{longrotatetable}
\begin{deluxetable*}{rrrrrrrrrrrrrrl}
	\tablecaption{Parameters of the 2443 star cluster  candidates identified by SHiP in this work. The position, parallax and 
	proper motion of each star cluster are calculated as the average value of all cluster members   with one sigma dispersion indicated.  
    The radius $r_\mathrm{max}$ is defined as the maximum distance of cluster members to the average position. For convenience, the 
	corresponding IDs in K13, CG18$+$19 and B19 are also listed ($-1$ means unmatched). All IDs 
	start with 0 and correspond to the line number in the catalog (not the MWSC 
	number as in K13). For CG18$+$19, IDs in the ranges  $0-1228$ and $1229-1274$ correspond to star
	clusters in \citet{CG18} and \citet{CG19}, respectively.   For
	convenience, the names of the matched star clusters in CG18$+$19 are 
	also presented. 
	Uncertainties of age and 
    metallicity are estimated as half a step in the isochrone table (see Sec.~\ref{sec:padova} for 
    more details).   These  are 6\% ($10^{\frac{1}{2}\Delta(\log t/\mathrm{yr})}$) for ages and 0.125 for metallicities. 
	A machine readable table is available online, {see data/cat\_all.txt in the github repository for a full version.}
	\label{tab:cat_all}}
	\tablewidth{0pt}
    \tabletypesize{\tiny}
	\tablehead{\colhead{FoF ID} & \colhead{$l$} & \colhead{$b$} & 
		\colhead{$r_\mathrm{max}$} & \colhead{$\varpi$} &
		\colhead{$\mu_\alpha\cos\delta$} & 
		\colhead{$\mu_{\delta}$} & \colhead{$n_\mathrm{tot}$} & 
		\colhead{$t_\mathrm{age}$} & \colhead{$Z$} & \colhead{Class} & 
		\colhead{K13 ID} & \colhead{CG18$+$19 ID} & \colhead{B19 ID} & 
		\colhead{Name}\\
		\colhead{} & \colhead{(deg)} & \colhead{(deg)} & \colhead{(deg)} & 
		\colhead{(mas)} & \colhead{(mas/yr)} & \colhead{(mas/yr)} & \colhead{} 
		& \colhead{(Gyr)} & \colhead{($\log\frac{Z}{\mathrm{Z}_\odot}$)} & 
		\colhead{} & \colhead{} & \colhead{} & \colhead{} & \colhead{}}
	\startdata
	0 & 186.181 $\pm$ 0.182 & -13.025 $\pm$ 0.173 & 0.817 & 0.549 $\pm$ 0.058 & 0.516 $\pm$ 0.252 & -0.888 $\pm$ 0.221 & 567 & 1.41 $\pm$ 0.08 & -0.250 & 1 &  352 &  687 & 5076 & NGC\_1817\\
1 & 184.719 $\pm$ 0.120 & -13.510 $\pm$ 0.077 & 0.470 & 0.359 $\pm$ 0.052 & 0.355 $\pm$ 0.224 & -2.500 $\pm$ 0.220 & 135 & 1.32 $\pm$ 0.08 & 0.000 & 1 &   -1 &   -1 & 5047 & -\\
2 & 226.034 $\pm$ 0.110 & -16.126 $\pm$ 0.116 & 0.461 & 0.252 $\pm$ 0.039 & -0.538 $\pm$ 0.237 & 1.975 $\pm$ 0.282 & 384 & 1.38 $\pm$ 0.08 & 0.250 & 1 &  576 &  708 & 6219 & NGC\_2204\\
3 & 239.472 $\pm$ 0.069 & -18.018 $\pm$ 0.062 & 0.331 & 0.257 $\pm$ 0.048 & -1.261 $\pm$ 0.151 & 5.493 $\pm$ 0.164 & 361 & 3.39 $\pm$ 0.20 & 0.000 & 1 &  625 &  713 & 6569 & NGC\_2243\\
4 & 259.574 $\pm$ 0.073 & -14.278 $\pm$ 0.089 & 0.384 & 0.258 $\pm$ 0.040 & -1.476 $\pm$ 0.200 & 2.740 $\pm$ 0.255 & 330 & 2.69 $\pm$ 0.16 & 0.000 & 1 &  917 &  651 & 7005 & Melotte\_66\\
5 & 292.316 $\pm$ 0.216 & -12.736 $\pm$ 0.129 & 0.559 & 0.506 $\pm$ 0.026 & -6.873 $\pm$ 0.177 & 1.425 $\pm$ 0.215 & 171 & 3.16 $\pm$ 0.19 & 0.000 & 1 &   -1 &   -1 &   -1 & -\\
6 & 325.553 $\pm$ 0.032 & -17.569 $\pm$ 0.035 & 0.118 & 0.242 $\pm$ 0.031 & -5.552 $\pm$ 0.208 & -4.683 $\pm$ 0.316 & 126 & 10.70 $\pm$ 0.64 & -2.000 & 3 & 2106 &   -1 & 9172 & -\\
7 & 0.069 $\pm$ 0.027 & -17.299 $\pm$ 0.031 & 0.133 & 0.280 $\pm$ 0.054 & 0.900 $\pm$ 0.379 & -2.391 $\pm$ 0.401 & 206 & 0.85 $\pm$ 0.05 & -2.000 & 3 & 2420 &   -1 &   10 & -\\
8 & 5.617 $\pm$ 0.034 & -14.071 $\pm$ 0.069 & 0.367 & 0.281 $\pm$ 0.063 & -2.958 $\pm$ 0.399 & -1.410 $\pm$ 0.429 & 176 & 0.0040 $\pm$ 0.0002 & 0.500 & 3 & 2412 &   -1 &  338 & -\\
9 & 8.793 $\pm$ 0.054 & -23.268 $\pm$ 0.054 & 0.265 & 0.341 $\pm$ 0.120 & -3.412 $\pm$ 0.404 & -9.270 $\pm$ 0.366 & 1798 & 12.90 $\pm$ 0.77 & -2.000 & 1 & 2506 &   -1 &  473 & -\\

	\enddata
\end{deluxetable*}
\end{longrotatetable}

\begin{deluxetable}{l|rrr|r}
\caption{Summary of the total number of cross matched star clusters  with K13, CG18$+$19 and B19. 
\label{tab:cross_match}}
\tablewidth{0pt}
\tablehead{\colhead{Catalog} & \multicolumn{3}{c}{Class} & \colhead{Total} \\
	\colhead{} & \colhead{1} & \colhead{2} & \colhead{3} & \colhead{}}
\startdata
K13		    &	439 &	72 	&	391 	&	902		\\
CG18$+$19	&	430	&	51 	&	233 	&	714 	\\
B19         &   444 &   62  &   370     &   876     \\
\enddata
\end{deluxetable}

\section{Star cluster candidates}\label{sec:sc_property}

\subsection{General properties}\label{sec:general_property}

\begin{figure}
\caption{The distributions of position, proper motion, parallax and the CMD of a typical Class 1 star cluster candidate   that cross-matched with CG18+19. 
The positions and proper motions of cluster members are plotted   as   offsets  to  the mean value of all members (the average position and proper motion of members are 
presented in the upper left corner of the corresponding panel). 
Blue dots represent  cluster members. Red crosses and lines  demonstrate the corresponding mean position, proper motion and parallax obtained from CG18$+$19. 
The black dotted curve in the CMD panel is  
the best-fitting  isochrone. The fitted  
parameters are presented in the same panel. A detailed description of these parameters is given in Sec.~\ref{sec:iso_parameter}.
 \label{fig:cls_1}}
\plotone{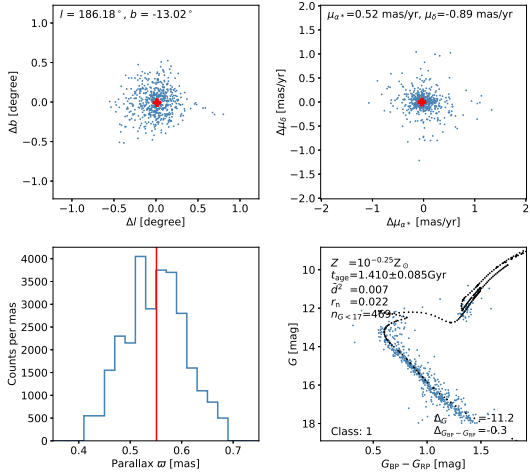}
\end{figure}

\begin{figure}
	\caption{The   distributions   of position, proper motion, parallax and the CMD of member stars of a typical Class 2 star cluster candidate. The symbols and colors are identified to those  in Fig.~\ref{fig:cls_1}.  	\label{fig:cls_2}}
	\plotone{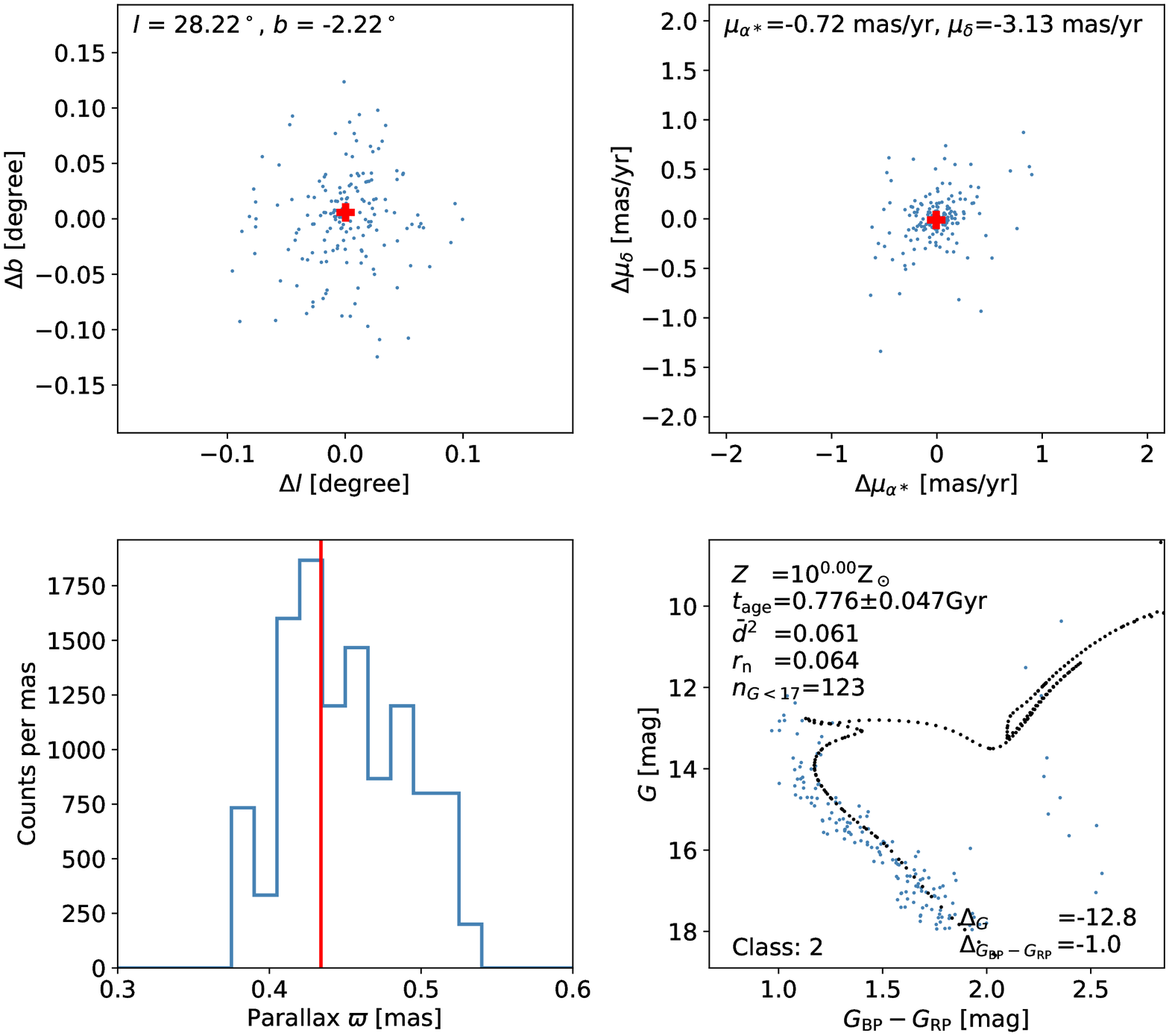}
\end{figure}

\begin{figure}
	\caption{The   distributions   of position, proper motion, parallax and the CMD of member stars of a typical Class 3 star cluster candidate. The symbols and colors are identified to those  in Fig.~\ref{fig:cls_1}.  
	\label{fig:cls_3}}
	\plotone{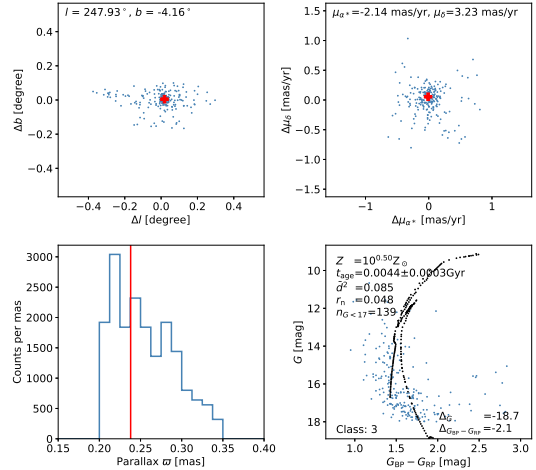}
\end{figure}

Among all 2443 star cluster candidates, those in Class 1 are likely genuine star clusters   which are mostly located within 4\,kpc . For these cluster candidates, a clear over-density is seen in the spatial and proper motion distributions of  cluster members (upper panels in  Fig.~\ref{fig:cls_1}). Member stars form a narrow MS in the CMD (bottom-right panel in Fig.~\ref{fig:cls_1}). On the contrary, star cluster candidates in Class 2 are more diffuse in distributions of position, proper motion and CMD (Fig.~\ref{fig:cls_2})\footnote{The 4-panel diagrams (Fig.~\ref{fig:cls_1}, \ref{fig:cls_2} and \ref{fig:cls_3}) 
 together with the member list of all 2443 candidates are available in the github repository of this work.}.   Note that the appearance of a few outliers may have  biased the fitting results. There is no standard criteria to automatically exclude stars from star cluster candidates. This can be done manually for individual clusters. 
For example, after removing the few faint red stars in Fig.~\ref{fig:cls_2} ($G_{\rm BP}-G_{\rm RP}>2$\,mag, and $G>14$\,mag), the fitted age changes from 776\,Myr to 380\,Myr, and the metallicity changes from log(Z/$\rm Z\odot$) = 0 to log(Z/$\rm Z\odot$)=$-0.75$. Therefore, a followup study of the Class 2 and Class 3 star cluster candidates may require visual inspection of their CMDs.  

  Although cluster members of Class 3 candidates are widely scattered in the CMD, without a clear MS (Fig.~\ref{fig:cls_3}),
there is still a clear and notable concentration of stars in the $(l, b)$ and proper motions (upper panels in Fig.~\ref{fig:cls_3}). The candidate shown in Fig.~\ref{fig:cls_3} is a cluster cross-matched with the CG18+19 catalog, which implies that there are genuine star clusters in Class 3 that deserve further investigation.

\subsection{Cross-match with previous catalogs}\label{sec:cross_matching}
We cross-match our identified star cluster candidates with three previously 
published catalogs: K13, CG18$+$19 and B19, so as to verify the reliability of our 
catalog. K13 is compiled by \citet{K13}, 
and contains 3006 objects   with  2MASS photometry; CG18$+$19 is a \Gaia DR2 based catalog, 
  which combines 1229 objects in \citet{CG18} and 46 objects in \citet{CG19}. B19 is a multi-band 
catalog with 10978 entries composed of Galactic star clusters, stellar associations and candidates in \citet{Bica2019}. 
To keep the same volume, star clusters with $|b| > 25^\circ$ 
are excluded, which   yields  2941 (K13), 1270 (CG18+19) and 10464 (B19) star clusters in the three catalogs for the following cross-matches, respectively. 

  A star cluster candidate in our catalog is  cross-matched after  comparison of their radii in the other catalogs.  
For our candidates, the radius $r_\mathrm{FoF}$ is 
defined as the maximum cluster-centric distance of members. We adopt the average position of members as the cluster center. For K13 we use $r2$, which is defined as the projected distance from the cluster centre where the surface density of cluster members is equal to the average surface density of the surrounding field \citep{K12}. For CG18$+$19, $r50$, the  radius which contains  half the number of
members, is adopted. For B19, we use the major axis of the cluster.  
  To obtain appropriate matching criteria, we test a variety of methods
using our catalog and CG18+19, such as, $d<(r_\mathrm{FoF} + r_\mathrm{REF})$,
$d<\max(r_\mathrm{FoF}, r_\mathrm{REF})$, etc. Here $d$ is the angular
distance between the centers of the two clusters.
$r_\mathrm{REF}$ is the radius of the star cluster provided in the
reference catalog. We further exclude artificial 
cross-matches that may occur when two clusters are located along nearly the same line of sight, by comparing their radii and angular
separations,   and the parallaxes .  After 
our investigation, a star cluster candidate in our catalog is regarded 
as cross-matched with  the cluster in the other catalogs when the angular
distance between their centers is smaller than any of their radii,
$d<\min(r_\mathrm{FoF}, r_\mathrm{REF})$. This cross-matching criterion 
yields the lowest   artificial  visual overlapping fraction and provides a 
reasonable matching number. 
430 star clusters candidates of Class 1 are cross-matched with CG18+19, 
3 clusters among these exhibit visual overlap\footnote{  The IDs of the three visual overlaps are 380, 2129 and 2281 in our catalog.  However, their cross-match with K13 or B19 still 
cannot be excluded (the parallax information provided by K13 does  agree not well 
with the \Gaia catalog. B19 provides no parallax information). };
for other criteria: $d<(r_\mathrm{FoF} + r_\mathrm{REF})$ and
$d<\max(r_\mathrm{FoF}, r_\mathrm{REF})$, artificial visual overlaps increase to 16 (456 cross-matched) and 13 (451   cross-matched ),
respectively.  
  In Tab.~\ref{tab:cat_all}, we present the corresponding ID
for each matched cluster.
 
The cross-matched results are summarized in Tab.~\ref{tab:cross_match}.
Overall, 902, 714 and 876 star cluster candidates are matched with K13, CG18$+$19 and B19, respectively.

 There is no difference in the metallicity and age distribution between the cross-matched and unmatched star cluster candidates. Generally, unmatched cluster candidates are much further away from the Sun than the cross-matched.  
Since CG18$+$19 is completely based on 
\Gaia DR2 data, the corresponding parameters are more consistent with our catalog. Among the 1270 star clusters ($|b|<25^\circ$) in CG18$+$19,   over half are identified by SHiP.   
We use CG18$+$19 to validate parameters of cross-matched star clusters derived from our pipeline. 

\begin{figure}
\caption{The discrepancy distributions of proper motions and parallaxes  for the matched star clusters, for our  identifications and those of CG18$+$19. Blue dashed histograms and red solid histograms  correspond to all matched clusters and 
those classified as Class 1, respectively. \label{fig:diff_CG18}}
\plotone{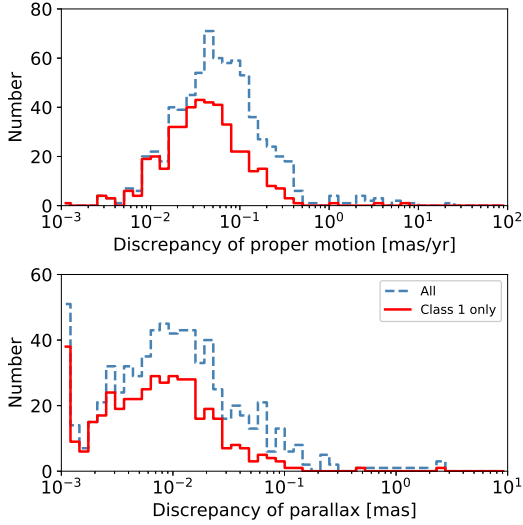}
\end{figure}

The mean values of position, proper motion and parallax from CG18$+$19 are plotted as red crosses and lines in Fig.~\ref{fig:cls_1}, \ref{fig:cls_2} and \ref{fig:cls_3}, which are consistent with the average value from our pipeline. Fig.~\ref{fig:diff_CG18} shows the discrepancy of proper motions and parallax of matched clusters compared to CG18$+$19 (blue dashed histograms).
  The peak around 0.1 mas/yr of the proper motions is smaller than the typical 
spread in the proper motions of the cluster stars ($\sim$1.0 mas/yr).  Similarly, the differences of the parallax are well within 0.1 mas, which is smaller than the uncertainty of the parallax at   $G \sim 18$\,mag  (0.2~mas).  Very few  star clusters have discrepancies in proper motion and parallax that are larger than 1 mas/yr and 10 mas,
respectively. 
Generally, the discrepancy distribution of the best fitted Class 1 candidates (red histograms in Fig.~\ref{fig:diff_CG18}) shows no difference for all matched samples (blue dashed histograms). 

\begin{figure}
\caption{The distributions of parallax (upper panel) and age (lower panel) of cross-matched star clusters with CG18$+$19. Red solid, cyan dotted and blue dashed histograms represent star cluster candidates in Class  1, 2, and 3, respectively. \label{fig:age_match_CG18}}
\plotone{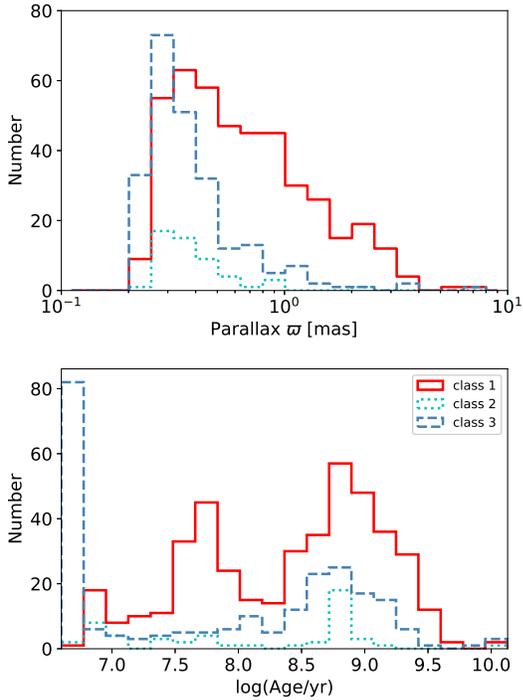}
\end{figure}

\begin{figure}
	\caption{The distributions of matched and new star cluster candidates of 
	different classes in the Galactic X-Y plane   with indication of spiral arms . The positions (mean position of
	cluster members) of identified cluster group
	candidates are also presented. \label{fig:Gal_XY}}
	\plotone{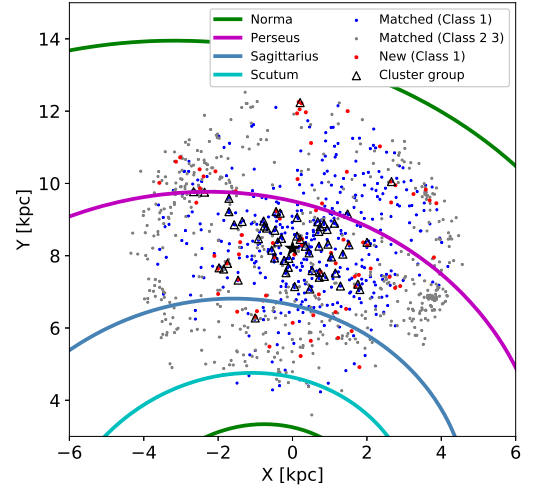}
\end{figure}

  Given the same cluster, its number of members brighter than $G$ = 17\,mag (parameter $n_{G<17}$) is larger when it is nearer. 
 Therefore, among the 714 matched star cluster candidates with CG18$+$19, Class 1 candidates are primarily star clusters within    4~kpc (larger value of parallax; see the upper panel in Fig.~\ref{fig:age_match_CG18}), and on average slightly closer than clusters in Class 2 and Class 3 (Fig.~\ref{fig:Gal_XY}).  
 
  At the same time,
the age distributions among the three classes (bottom panel in Fig.\ref{fig:age_match_CG18}) are distinct. Class 1 clusters show two populations. The young population has a broad peak around log(age/yr)$\sim$7.8 and the old population at log(age/yr)$\sim$8.8. Class 2 only has a dominant old population at log(age)/yr$\sim$8.8. 
Similar to Class 1 and 2,  there is also an old population at log(age/yr)$\sim$8.8 in Class 3. Additionally, Class 3 contains a very young population at log(age/yr)$<6.8$, which does not exist in other two classes. This is due to our age cut at 5\,Myr for Class 1 and 2.   On the other hand, studies of solar neighborhood star clusters did find an excess of young star clusters with ages~$\le$~9\,Myr \citep{{Bonatto2011}}. Therefore, the young cluster candidates in Class 3 deserve further investigation.  
Old star clusters with  log(age/yr)$\sim$8.8 have a prominent MS turn-off in the CMD, enabling reliable isochrone fitting. 
   At the age of log(age/yr)$\sim$7.8 (Class 1),  
 the radiation of young massive stars has already dispersed the parent molecular clouds. Therefore, such clusters are free from differential reddening and have a relatively narrow MS that enables reliable isochrone fitting with our pipeline.  In the  discussion below, we will focus on the most probable cluster candidates in Class 1.  

\subsection{New star cluster candidates in Class 1}
 
 \begin{deluxetable*}{rrrrrrrrrc}
	\caption{The 76 new star cluster  candidates classified as Class 1.
	A machine readable table is available online; {see  data/cat\_new.txt in the github repository for a full version.}
		\label{tab:cat_new}}
	\tablewidth{0pt}
	\tablehead{\colhead{FoF ID} & \colhead{$l$} & \colhead{$b$} & 
	\colhead{$r$} & \colhead{$\varpi$} & \colhead{$\mu_\alpha\cos\delta$} & 
	\colhead{$\mu_{\delta}$} & \colhead{$n_\mathrm{tot}$} & 
	\colhead{$t_\mathrm{age}$} & \colhead{$Z$} \\
	\colhead{} & \colhead{(deg)} & \colhead{(deg)} & \colhead{(deg)} & 
	\colhead{(mas)} & \colhead{(mas/yr)} & \colhead{(mas/yr)} & \colhead{} & 
	\colhead{(Gyr)} & \colhead{($\log\frac{Z}{\mathrm{Z}_\odot}$)}}
	\startdata
	5 & 292.316 $\pm$ 0.216 & -12.736 $\pm$ 0.129 & 0.559 & 0.506 $\pm$ 0.026 & -6.873 $\pm$ 0.177 & 1.425 $\pm$ 0.215 & 171 & 3.16 $\pm$ 0.19 & 0.000\\
58 & 264.972 $\pm$ 0.232 & -2.881 $\pm$ 0.093 & 0.768 & 0.524 $\pm$ 0.034 & -5.813 $\pm$ 0.286 & 5.063 $\pm$ 0.251 & 482 & 0.02 $\pm$ 0.00 & 0.250\\
145 & 343.191 $\pm$ 0.068 & -2.218 $\pm$ 0.048 & 0.208 & 0.518 $\pm$ 0.049 & -2.103 $\pm$ 0.251 & -5.424 $\pm$ 0.230 & 264 & 0.98 $\pm$ 0.06 & 0.250\\
198 & 234.945 $\pm$ 0.107 & -1.276 $\pm$ 0.039 & 0.347 & 0.310 $\pm$ 0.041 & -1.972 $\pm$ 0.534 & 2.311 $\pm$ 0.589 & 147 & 0.54 $\pm$ 0.03 & -0.750\\
273 & 333.632 $\pm$ 0.028 & -0.346 $\pm$ 0.030 & 0.101 & 0.362 $\pm$ 0.036 & -2.346 $\pm$ 0.365 & -4.109 $\pm$ 0.289 & 97 & 1.48 $\pm$ 0.09 & -0.750\\
282 & 355.798 $\pm$ 0.032 & -1.447 $\pm$ 0.031 & 0.095 & 0.335 $\pm$ 0.025 & -0.027 $\pm$ 0.244 & -0.994 $\pm$ 0.299 & 75 & 0.69 $\pm$ 0.04 & 0.500\\
321 & 57.818 $\pm$ 0.047 & -1.706 $\pm$ 0.043 & 0.151 & 0.423 $\pm$ 0.049 & -0.360 $\pm$ 0.187 & -3.504 $\pm$ 0.197 & 132 & 0.68 $\pm$ 0.04 & 0.500\\
386 & 250.132 $\pm$ 0.135 & 0.927 $\pm$ 0.079 & 0.315 & 0.255 $\pm$ 0.026 & -2.534 $\pm$ 0.378 & 3.076 $\pm$ 0.502 & 193 & 0.58 $\pm$ 0.03 & 0.250\\
403 & 286.748 $\pm$ 0.048 & 0.721 $\pm$ 0.038 & 0.172 & 0.367 $\pm$ 0.031 & -6.939 $\pm$ 0.345 & 3.019 $\pm$ 0.255 & 164 & 0.0060 $\pm$ 0.0004 & 0.500\\
589 & 237.710 $\pm$ 0.067 & 5.499 $\pm$ 0.040 & 0.212 & 0.304 $\pm$ 0.033 & -1.398 $\pm$ 0.172 & 0.536 $\pm$ 0.174 & 87 & 0.76 $\pm$ 0.05 & -0.750\\

	\enddata
\end{deluxetable*}

\begin{figure}
\caption{The distribution of star cluster candidates classified as Class 1 in 
Galactic coordinates. Blue crosses and red dots correspond to the cross-matched and newly identified star cluster candidates, respectively. \label{fig:mollweide}}
\plotone{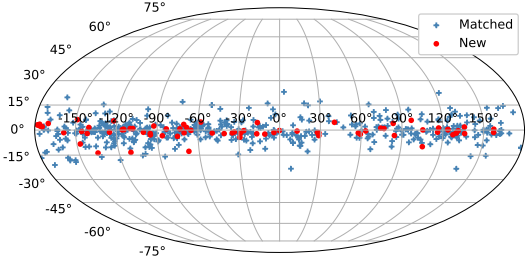}
\end{figure}

\begin{figure}
	\caption{The distributions of the Galactic longitude ($l$) and latitude ($b$), parallax ($\varpi$) and age (derived using  isochrone fitting) of all matched  (blue histograms), matched Class 1 (blue dotted histograms)   with K13, CG18+19 and B19, and new star cluster candidates (red dashed histograms) from Class 1. \label{fig:schist}}
	\plotone{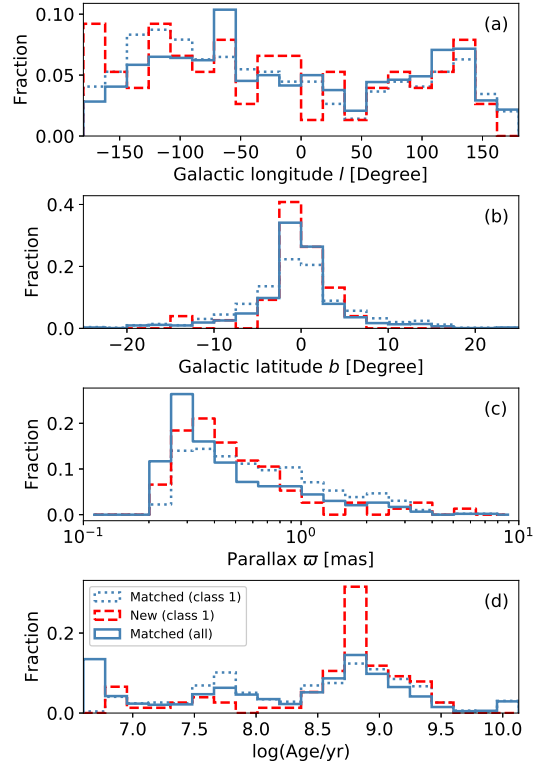}
\end{figure}

A total of 76 star cluster candidates classified as Class 1  are not present in any 
 of the three earlier  catalogs, which accounts for 13.6\% of the total number of clusters in Class 1 (see Tab.~\ref{tab:cat_new} for a full list).
 These previously uncataloged candidates are very likely genuine star clusters.    Fig.~\ref{fig:Gal_XY}  and Fig.~\ref{fig:mollweide} demonstrate the spatial distribution of cross-matched clusters in Class 1 (blue dots and crosses) and new star cluster candidates (red dots) in Galactic coordinates,  which indicates a disk concentration for both groups,   spreading out in the  inter-arm regions (Fig.~\ref{fig:Gal_XY}).   It is necessary to carry
out a systematic study to investigate the general properties of new cluster candidates, in particular 
their differences from known star clusters cross-matched with the catalogs of K13, CG18+19, and B19. As can be seen in Fig.~\ref{fig:schist} (panels a, b, and c), there is no major difference between the new cluster candidates and the cross-matched ones (Class 1 and all classes) in Galactic longitude, latitude and parallax.
All of these are 
distributed mainly within $|b|\sim10^\circ$. 
Both the cross-matched and new cluster candidates show an old population peaked at log(age/yr)$\sim$8.8. However, the excess of old star clusters is more prominent among the newly identified  candidates.


\subsection{  Candidates of cluster groups }

\begin{deluxetable}{l|rrrr}
\caption{The 56 cluster groups identified in our catalog, see data/group
folder in the github repository for a full version.\label{tab:cat_group}}
\tablewidth{0pt}
\tablehead{\colhead{Group ID} & \colhead{FoF ID} & \colhead{$l$} &
\colhead{$b$} & \colhead{$\varpi$}\\
\colhead{} & \colhead{} & \colhead{(deg)} & \colhead{(deg)} &
\colhead{(mas)}}
\startdata
0 & 30 & 236.057 & -4.639 & 0.569\\
   & 51 & 238.210 & -3.327 & 0.563\\
\hline
1 & 2088 & 265.114 & -2.573 & 0.506\\
   & 58 & 264.972 & -2.881 & 0.524\\
   & 105 & 264.190 & -1.572 & 0.486\\
   & 2087 & 266.363 & -1.917 & 0.476\\
\hline
2 & 1235 & 27.797 & -1.492 & 0.469\\
   & 155 & 27.306 & -2.778 & 0.453\\
\hline
3 & 171 & 120.315 & -2.547 & 0.507\\
   & 894 & 121.982 & -2.676 & 0.496\\
   & 680 & 120.136 & -4.820 & 0.502\\
\hline
4 & 936 & 235.366 & 0.153 & 0.307\\
   & 198 & 234.945 & -1.276 & 0.310\\
\hline
5 & 422 & 299.728 & 0.844 & 0.493\\
   & 251 & 299.034 & -0.348 & 0.500\\
\hline
6 & 358 & 120.778 & -0.953 & 0.320\\
   & 1321 & 120.355 & -0.371 & 0.327\\
\hline
7 & 425 & 303.206 & 2.497 & 0.469\\
   & 2153 & 300.951 & 1.228 & 0.466\\
\hline
8 & 496 & 73.236 & 1.263 & 0.510\\
   & 493 & 72.653 & 2.067 & 0.522\\
\hline
9 & 1819 & 112.741 & 0.883 & 0.581\\
   & 526 & 112.808 & 0.430 & 0.595\\

\enddata
\end{deluxetable}

    Studies of
clustering among OCs provide keys to understanding star formation in the Galactic disk and the subsequent dynamical evolution of star clusters.  
We carry out   a  search for OC groups  using the FoF method amongst the clusters in our catalog (Table 1). The   procedure to identify these OC groups  is summarized as follows: 
\begin{itemize}
    \item   Since Class 1 cluster candidates are mostly at nearby clusters with good photometry, we 
    only search for groups among the  Class 1 candidates in our catalog. 
    \item We convert the coordinates and parallaxes of selected star cluster
    candidates to 3-D Cartesian coordinates. 
    \item We carry out a  standard FoF group  {identification} with a linking length of 100~pc   \citep{Conrad2017} .
\end{itemize}
  \citet{Conrad2017} have carried out FoF search for cluster groups in the 6-D parameter space (3-D position and 3-D velocity).  
 Unlike in \citet{Conrad2017}, radial velocities are not taken into account in our current 
catalog, our search of cluster groups is based on the star clusters' 3-D
positions only. In total, we identify 152 star cluster candidates 
distributed amongst 56 cluster groups (See Tab.~\ref{tab:cat_group}). We plot our   group candidates  in the Galactic X-Y plane as black triangles in Fig.~\ref{fig:Gal_XY}.   These group candidates are mainly concentrated in the Solar neighborhood, within distances 2~kpc, between the Perseus and Sagittarius arms. Some group candidates only contain two star clusters, and may be candidates of binary cluster \citep[see, e.g.,][]{Priyatikanto2016}. However, kinematic data including a sufficient number of stars with radial velocities are required to confirm the dynamical status of these OC group candidates.

\section{Summary}\label{sec:summary}
In this work, we identify star clusters   in  the Milky Way disk from \Gaia DR2 data with the Star cluster Hunting Pipeline, SHiP. Our main  results are summarized as follows:
\begin{itemize}
    \item Star clusters are  identified with our FoF  
    cluster finder in the 5-D parameter space composed of the position, parallax
    and proper motion. They are further verified with an isochrone 
    fitting  {scheme}. We classify star cluster candidates into 3 classes based 
    on four parameters: $\bar{d^2}$, $r_\mathrm{n}$, $n_{G<17}$, $t_\mathrm{age}$. 
    SHiP is designed in a highly parallel and automated 
    way, which makes it possible to identify  star clusters in \Gaia 
    DR2 without any prior information.
    \item In total, 2443 star cluster candidates are detected, which are classified into 3 classes (with 569, 127 
    and 1747 cluster candidates in Class 1, 2 and 3, respectively). In our classification scheme, those 
    classified as Class 1 are likely star clusters, showing a narrow MS in the CMD and concentration in spatial and proper motion distributions.   Class 1 and 2 clusters are located at distances mainly within 4~kpc, due to the imposed constraint $n_{G<17}$. 
    \item   A total of 902, 714 and 876 star clusters candidates in our catalog are cross-matched with K13, CG18$+$19 and B19. respectively. Our star cluster catalog (Tab.~\ref{tab:cat_all}) is in a good agreement  with previously published 
    catalogs. The discrepancies of proper motion and 
    parallax of the matched clusters with CG18$+$19 are well within observational uncertainties.
    \item A total of 76 new star cluster candidates are detected in Class 1. These were not present in any of the earlier three catalogs. These new star cluster candidates are concentrated towards the very thin disk  ($|b|<5^\circ$). The majority of these are clusters older than log(age/yr)~=~8.0 with prominent narrow MSs. 
    \item   56 candidates of star cluster group are identified by the FoF group finder based on star clusters' 3-D positions. They are distributed within 2~kpc of the Sun, between the Perseus and Sagittarius spiral arms. Further investigations are necessary to confirm the nature of these group candidates.  
\end{itemize}
The new star cluster candidates found in this work suggest that the current 
 star cluster catalogs in the Milky Way are still incomplete. 
All the necessary materials, including the pipeline, catalog, 4-panel figures (including position, proper motion, parallax and CMD), member list of each  individual star cluster 
candidate,   star-cluster-group candidates,  are available in the github repository: \href{https://github.com/liulei/gaia_ship}{\seqsplit{https://github.com/liulei/gaia\_ship}}.

\acknowledgments
 The authors are grateful for the financial support from the National Natural Science Foundation of China through grants, No. 11903067, 11673032, 11503015 and U1938114. X.Y.P. expresses gratitude for support from the Research Development Fund of Xi'an Jiaotong Liverpool University (RDF-18-02-32). This study was supported by Sonderforschungsbereich SFB 881 “The Milky Way System”  (sub-project B5)  of  the  German Research  Foundation (DFG). We thank the referee for his/her suggestions   which  greatly helped    to improve the quality of this paper.  We are also grateful to Dr. Thijs Kouwenhoven for in-depth discussions.

\end{document}